\documentstyle [12pt,epsfig]{article} 
\makeatletter

\@addtoreset{equation}{section}
\makeatother

\newcommand{\ba}{\begin{eqnarray}}
\newcommand{\ea}{\end{eqnarray}}

\begin{document}
\newcommand{\BS}{\bigskip}
\newcommand{\SECTION}[1]{\BS{\large\section{\bf #1}}}
\newcommand{\SUBSECTION}[1]{\BS{\large\subsection{\bf #1}}}
\newcommand{\SUBSUBSECTION}[1]{\BS{\large\subsubsection{\bf #1}}}
\begin{titlepage}
\begin{center}
\vspace*{2cm}
{\large \bf Einstein and Planck on mass-energy equivalence in 1905-06:
          a modern perspective}  
\vspace*{1.5cm}
\end{center}
\begin{center}
{\bf J.H.Field }
\end{center}
\begin{center}
{ 
D\'{e}partement de Physique Nucl\'{e}aire et Corpusculaire
 Universit\'{e} de Gen\`{e}ve . 24, quai Ernest-Ansermet
 CH-1211 Gen\`{e}ve 4.
}
\newline
\newline
   E-mail: john.field@cern.ch
\end{center}
\vspace*{2cm}
\begin{abstract}
    Einstein's theoretical analysis of mass-energy equivalence, already, at the time, experimentally
   evident in radioactive decays, in two papers published in 1905,
   as well as Planck's introduction, in 1906, of the concepts of relativistic momentum, and,
   by invoking Hamilton's Principle, relativistic energy, are reviewed and discussed.
    Claims in the
   literature that Einstein's analysis was flawed, lacked generality, or was not
   rigorous, are rebutted.  

\vspace*{1cm}
\end{abstract}
\end{titlepage}
 
\SECTION{\bf{Introduction}}
\par Laws of physics are expressed as mathematical equations, so in order to discover a new
  law the corresponding equation must first be written down. However, the meaning
  of the law is only made manifest when the precise connection between the symbols in the equation
  and measureable physical phenomena is made clear. The discoverer of the equation is therefore
  not necessarily the same person who elucidates the meaning of the corresponding physical law.
  Once the latter is established it can be expressed either as an equation, where the operational
   meaning of all the symbols is quite clear, or alternatively, in words, in which case it
   can be better understood by less mathematically literate readers. It will be shown below that 
   in Einstein's seminal work on mass-energy equivalence, published in 1905, both of the above possiblities
   for the formulation of a new physical law were realised. In fact it was aleady manifest in
   an equation  in the orginal special relativity paper of June 1905~\cite{Ein1} when this equation
   is correctly interpreted physically, but this interpretation was not given by Einstein. In contrast,
   in the September 1905 paper~\cite{Ein2}, where the equivalence of mass and energy was clearly stated
   verbally, the corresponding equation was not written down. 
   \par  In the present paper, Einstein's September 1905 paper
   `Does the inertia of a body depend on its energy content'~\cite{Ein2} and Planck's 1906
    paper on relativistic kinematics~\cite{Planck}
    will be reviewed from three different perspectives: (i) Einstein's and Planck's 
    original arguments. (ii)
    Arguments Einstein or Planck could have used, given the state of knowledge in 1905-06.
    (iii) A review of the related literature
    critical of Ref.~\cite{Ein2} which claimed that Einstein's arguments were flawed or lacked
    generality or rigour. This paper is organised as follows: in the two following sections
    the arguments of Ref.~\cite{Ein2} and Ref.~\cite{Planck}, respectively, are reviewed. In Section 4
    equations given in Einstein's June 1905 paper~\cite{Ein1} are discussed in relation to the conclusions
     of his September 1905 paper~\cite{Ein2} and the related work of Planck. Section 5 deals with criticisms
     of Ref.~\cite{Ein2}
     by Planck~\cite{Planck1907}, Ives~\cite{Ives}, Ohanian~\cite{Ohanian} and Hecht~\cite{Hecht}. 
     Conclusions are given in Section 6 . 
     \par Throughout this paper the concept of `mass' (denoted by $m$) refers only to the Lorentz-invariant
      quantity `rest mass' that is proportional to the energy in the static limit. For a critical
       discussion of the concept of velocity-dependent mass see Ref.~\cite{Okun} and references therein.
    \par Discussion of Einstein's work on  mass-energy equivalence after 1905 is beyond the scope
      of the present paper. See Refs.~\cite{Fadner, Hecht, Hecht1} for references to this later work 
     and the related literature.
 
\SECTION{\bf{Einstein's September 1905 paper on mass-energy equivalence}}
    As Einstein's original notation is somewhat cumbersome, a more modern one, similar
   to that of Ref.~\cite{Hecht}, will be used for the analysis of Einstein's
   gedanken experiment.
 A process in which a body
     radiates `plane waves of light' of equal energy content, in opposite directions, is
     considered both in the frame in which the body is at rest and one in which it moves with constant
    speed $v$. Since, in the rest frame of the body, no momentum is carried away by the light,
    then by Newton's Second Law no force acts on the body in this frame.
     The velocity of the body, $f$, produced
    by the radiation process from the body $i$, is then the same as that of $i$, in any frame.
    Einstein assumes
     energy conservation in both frames:
     \begin{eqnarray}
      E_i & = & E_f +E(L) = E_f+\Delta E_i, \\                                                
  E'_i & = & E'_f +E'(L) = E'_f+\Delta E'_i
      \end{eqnarray}
      where primed quantities refer to the frame in which the objects are in motion 
     and $\Delta E_i$, $\Delta E'_i$
      are the absolute values of the changes in the energy of $i$ due to the radiation, which are
       equal to the energies $E(L)$, $E'(L)$, respectively, of the radiated light.
      Note that, already at this stage, Einstein has tacitly introduced in (2.1) and (2.2) 
      the concept of the `rest energies' $E_i$ and $E_f$ of the objects $i$ and $f$ as well
      as the `total energies'  $E'_i$ and $E'_f$ of the same objects in motion.
    Using the formula from \S 8 of
       Einstein's earlier
       1905 relativity paper~\cite{Ein1} to transform the energy of a `light complex' (identified with the 
      total energy of the two `plane waves' of light)
      between the two inertial frames gives:
      \begin{equation}
       E'(L) = \frac{E(L)}{\sqrt{1-\beta^2}} 
      \end{equation}
       where $\beta \equiv v/c$. Subtracting (2.1) from (2.2) and using (2.3) gives
         \begin{equation}
         E'_i- E_i-(E'_f-E_f) = E(L)\left[\frac{1}{\sqrt{1-\beta^2}}-1\right].  
        \end{equation}                                                            
      The kinetic energies of the bodies $K_i$, $K_f$ are introduced as:
         \begin{eqnarray} 
            K_i & \equiv & E'_i-E_i-C, \\
            K_f & \equiv & E'_f-E_f-C
         \end{eqnarray}
          where $C$ is a constant, assumed to be the same for the bodies $i$ and $f$.
            Combining (2.4),(2.5) and (2.6) and retaining only O($\beta^2$) terms on the right side of
          (2.4) gives 
            \begin{equation}
              K_i -   K_f = \frac{1}{2}\frac{E(L)v^2}{c^2} +{\rm O}(\beta^4).
             \end{equation}    
            This is the last equation in Ref.~\cite{Ein2}. Einstein then states, \`{a} propos of
             this equation:\footnote{ The symbols use to denote energies in Ref.~\cite{Ein2} are replaced
             by the corresponding symbols used in the present paper.}

             \begin{quotation}
           {\tt From this equation it follows directly that :---}
             \par {\it If a body gives off the energy E(L) in the form of radiation, its 
               mass diminishes by $ E(L)/c^2$.} {\tt The fact that the energy withdrawn from the
               body becomes energy of radiation evidently makes no difference so that we are
              lead to the more general conclusion that
             \par The mass of a body is a measure of its energy content; if the \newline energy changes
                  by $\Delta E$, the mass changes in the same sense by \newline $\Delta E/[9 \times 10^{20}]$,
                the energy being measured in ergs, and the mass in grammes.
              \par It is not impossible that with bodies whose energy-content is \newline variable 
                to a high degree (e.g. with radium salts) the theory may be successfully put
              to the test.
              \par If the theory corresponds to the facts, radiation conveys inertia between the
                        emitting and absorbing bodies.} (Einstein's italics)
               \end{quotation}    
               The content of Einstein's italicised statement and the following sentence, written 
               as an equation is:
                \begin{equation}
                    m_i-m_f =\frac{E(L)}{c^2} = \frac{\Delta E_i}{c^2}.
                \end{equation} 
                The last three paragraphs of Ref.~\cite{Ein2}, quoted above, show that Einstein had not 
                only discovered Eq.~(2.8) but also understood, very clearly, the physical
                significance of the equation.
                 \par In order to derive (2.8) from (2.7) it is necessary to assume the Newtonian
                formula for kinetic energy: $K(v) = (1/2)mv^2$. It must be assumed that Einstein did this
                 tacitly. In may appear, because of this assumption, that the relation (2.8) is only an 
                 approximate one, valid when $v \ll c$. However, as pointed out by Stachel and Torreti
                  ~\cite{ST}, by defining the rest mass of an object as:
                    \begin{equation}
                    m \equiv \left[Lim~v \rightarrow 0\right] \frac{K(v)}{v^2/2}
                 \end{equation}
                   and assuming that Newtonian mechanics is the correct $v \rightarrow 0$ limit
                of relativistic kinematics, enables (2.7) to yield the relation:
               \begin{equation}
              \left[Lim~v \rightarrow 0\right]\left\{ \frac{K_i}{v^2/2}- \frac{K_f}{v^2/2}
                 \right\} =  \left[Lim~v \rightarrow 0\right]
                 \left\{\frac{\Delta E_i}{c^2} +{\rm O}(\beta^2)\right\}
              \end{equation}
                  from which, on using (2.9), (2.8) immediately follows. Einstein's 
            final (verbally presented) equation
                  (2.8) is therefore an exact (velocity independent) relation.
                 Setting $m_f = 0$, $\Delta E_i = E_i$ in (2.8) gives:
                     \begin{equation}
                    E_i(\beta = 0) = m_i c^2.
               \end{equation}
               This is indeed $E_0 = m c^2$. Einstein's final verbal statement of the equivalence of mass
               and energy, expressed by Eq.~(2.8), is a necessary consequence of his premises which are: (A) conservation
               of energy, together with the tacit introduction in Eqs.~(2.1) and (2.2) of the
                concept of `total energies' of objects either at rest or in motion (B) the transformation law for the energy of radiation  and
              (C) Newtonian mechanics as
            the low velocity limit of relativistic mechanics. Whether a philosopher would consider
              this derivation `rigorous' is perhaps an open question but I submit, in agreement
                with Stachel and Torreti~\cite{ST} and Fadner~\cite{Fadner}, and contrary to Planck~\cite{Planck},
        Ives~\cite{Ives} and 
    recent assertions of Hecht~\cite{Hecht} and Ohanian~\cite{Ohanian}, that most physicists
              would. Further discussion of this controversial point is found in Section 6 below.
\SECTION{\bf{Planck's 1906 derivation of relativistic energy and momentum}}
     In Ref.~\cite{Planck} Planck unconventionally denoted the Lagrangian function  by $H$ and the
      Hamiltonian function by $L$. In the present discussion, to be in accordance
     with modern convention, this nomenclature will be inverted, and
      the scaled velocity in 3-vector notation $\vec{\beta} =(1/c)d\vec{s}/dt$ where
    $\vec{s}$ is the spatial displacement
      of an object, of rest mass $m$, in motion in free space, employed throughout. 
     The essential initial ansatz\footnote{In Ref.~\cite{Planck} the formula (3.1)
 was actually obtained by manipulation of a modified Lorentz force equation, using reasoning that
   is unclear to the present author.} of Ref.~\cite{Planck} is essentially a relativistic statement of
     Newton's Second Law of mechanics containing a definition of relativistic momentum:
     \begin{equation}
       \frac{d \vec{p}}{dt} \equiv \frac{d }{dt}\left(\frac{mc\vec{\beta}}{\sqrt{1-\beta^2}}\right)
   = \vec{F} 
     \end{equation}
      where $\vec{F}$ is a force that Planck considered to be produced by electric and magnetic fields,
      although
      this is inessential in the subsequent derivation of the formulas for relativistic energy.
       From the Lagrange equation, a necessary consequence of Hamilton's Principle~\cite{HP}:
         \begin{equation}
        p = \frac{1}{c}\frac{\partial L}{\partial \beta} = \frac{mc\vec{\beta}}{\sqrt{1-\beta^2}}
        \end{equation}
       and (3.1) Planck derived the Lagrangian, $L$, for a free particle:
          \begin{equation}
          L = -mc^2\sqrt{1-\beta^2} + C
        \end{equation}
          where $C$ is an arbitary constant. The Hamiltonian, $H$, is constructed from the
         momentum $\vec{p}$ and the Lagrangian according to the relation~\cite{Goldstein}
         \begin{eqnarray}
          H = E & = & c \vec{\beta}\cdot \vec{p} - L \nonumber \\
                & = & mc^2\left[\frac{\beta^2}{\sqrt{1-\beta^2}}+ \sqrt{1-\beta^2}\right]-C
           \nonumber \\
             & = & \frac{mc^2}{\sqrt{1-\beta^2}} -C. 
          \end{eqnarray}
   The quantity $E$ is called by Planck `lebendige Kraft'. If this is interpreted as `kinetic energy'
   then the constant $C$ in (3.4) takes the value $mc^2$ and, writing the conventional symbol $T$
   for kinetic energy, (3.4) reduces to the formula:
 \begin{equation}
        T = mc^2\left[\frac{1}{\sqrt{1-\beta^2}}-1\right] 
 \end{equation}
  as derived in Einstein's first special relativity paper~\cite{Ein1}. Alternatively
  choosing $C= 0$ in (3.4) gives the total relativistic energy:
 \begin{equation}
        E = \frac{mc^2}{\sqrt{1-\beta^2}}.
 \end{equation}
   Eqs.~(3.5) and (3.6) show that the total and kinetic relativistic energies
   are related according to:
  \begin{equation}
           E = T + mc^2.
 \end{equation}
    The famous formula $E_0\equiv E(\beta = 0) = mc^2$ is an immediate consequence of (3.7). 
    Planck gave at first however only the formula (3.4) and there was no discussion 
    of the value or physical significance of the constant $C$,
    or of any distinction between total, kinetic and rest relativistic
    energies. However (3.6) can be written as:
      \begin{equation}
        E = mc^2\left[\frac{1}{1-\beta^2}\right]^{\frac{1}{2}} 
          =  mc^2\left[1+\frac{\beta^2}{1-\beta^2}\right]^{\frac{1}{2}}    
          =  mc^2\left[1+ \frac{p^2}{m^2c^4}\right]^{\frac{1}{2}}
      \end{equation}
     where in the last member the definition in (3.1) of relativistic momentum, $p$, has been used.
     It also follows from (3.7) and (3.8) that:
     \begin{equation}
        T  =  mc^2\left[1+ \frac{p^2}{m^2c^4}\right]^{\frac{1}{2}}-mc^2.
      \end{equation}
      Planck did give Eq.~(3.8) (with an additional additive constant, $C_1$, on the right side)
      in Ref.~\cite{Planck}. 
      The meaning of Planck's quantity $L$ ($H$ in the notation of the present paper) then depends on the value of this constant $C_1$: $L \equiv E$,  $C_1=0$ 
      or  $L \equiv T$,  $C_1=-mc^2$. Eq.~(3.8), Planck's formula with
      $C_1=0$, when transposed, is nothing else than Lorentz
     invariant relation between the relativistic energy, relativistic
      momentum and the rest
     mass
    of a moving ponderable object in free space:
       \begin{equation} 
         E^2- p^2c^2 = m^2 c^4.
    \end{equation}
 
\SECTION{\bf{Further thoughts on Einstein's original 1905 special relativity paper}}
 Now for point (ii), the consideration of further arguments, concerning mass-energy equivalence,
    that Einstein could have used,
             given his knowledge of the contents of his first 1905 relativity paper~\cite{Ein1}.
                   In this paper Einstein derived the kinetic energy, $K$, of a ponderable object
                  in terms of its rest mass, $m$, and velocity, $v = \beta c$, as: 
      \begin{equation}
            K(m,\beta) = mc^2\left[\frac{1}{\sqrt{1-\beta^2}}-1\right]. 
       \end{equation}
          Clearly $ K(m,\beta = 0) = 0$.  Writing explicitly the velocity dependence of the energies in
         (2.5) and (2.6):
        \begin{equation}
          K_{i,f}(m,\beta) = E'_{i,f}(\beta) - E_{i,f}(0) -C.
        \end{equation}
         Since from (4.1) $K_{i,f}(m,0) = 0$ and $E'_{i,f}(0) =  E_{i,f}(0)$ Einstein's constant $C$ in
         (2.5) and (2.6) must vanish. Combining (2.5) and (2.6) (with or without $C = 0$) with (2.4) gives:
        \begin{equation}
          K_i-K_f = E(L)\left[\frac{1}{\sqrt{1-\beta^2}}-1\right]  
        \end{equation}  
         which is the exact relation, of which (2.7) is the O($\beta^2$) approximation.
          Combining (4.1) and (4.3) gives:
        \begin{equation}
          K_i-K_f = (m_i-m_f)c^2\left[\frac{1}{\sqrt{1-\beta^2}}-1\right]
            = E(L)\left[\frac{1}{\sqrt{1-\beta^2}}-1\right] 
           \end{equation}  
          from which Eq.~(2.8) follows as an exact (to all orders in $\beta$) relation, without any necessity
       to consider the $\beta \rightarrow 0$ limit, as in the derivation of Eq.~(2.8) in Section 2 above.
            \par It is now interesting to recall that the factor $[1/\sqrt{1-\beta^2}-1]$ 
            that cancels from both sides of the last member in (4.4) is derived in
          different and independent ways in Eq.~(4.1) and (4.3). In Eq.~(4.3) it follows from
          the transformation of the energy of `plane waves of light'. The kinetic energy
            in (4.1) is instead found by calculating the work, $W$, done on an electron, 
            initially at rest, acted on by a constant electric field. For this the
             relativistic generalisation of Newton's Second Law given in \S 10 of
             Ref.~\cite{Ein1} is used:
      \begin{equation}
        m \gamma^3 \frac{d^2 x}{d t^2} =  m \gamma^3 \frac{dv}{dt} = e E_x = F = {\rm const}
        \end{equation}
          where $\gamma \equiv 1/\sqrt{1-\beta^2}$ to give
  \begin{equation}
   K = W = \int dW = \int F dx = m\int_0^v\gamma^3\frac{dx}{dt} dv = m\int_0^v\gamma^3 v dv
                   = mc^2(\gamma-1).
     \end{equation}
        Although an electrostatic force is considered in (4.5) it is clear that the relation
       (4.6) is of complete generality, independent of the nature of the constant force $F$. Thus
        this relation holds, not only for electrons, but for any ponderable object of rest mass $m$ subjected
        to a constant force. 
        Transposing (4.6) gives:
  \begin{equation}
      K(m,\beta)+m c^2 \equiv E(m,\beta) = \gamma m c^2.
   \end{equation}
    The relation $E(m,0) = m c^2$ is evidently a direct consequence of (4.7), so that a special
     thought experiment such as that discussed by Einstein in Ref.~\cite{Ein2} was not necessary
     to derive it. However, this thought experiment is of great importance as {\it the first example of a theoretical
     analysis of the 
     transformation of part of the mass of an object into the energies of different physical
        objects} 
      ---Einstein's `plane waves of light' but what we would today understand, in a realistic
       realisation of the thought experiment, as a pair of photons of equal energy and 
       opposite momenta. 
      It can be argued that the fact that this
    phenomenon ---modification or destruction and creation of particles--- 
    occurs in nature, is the 
     single most important physical discovery of the 20th Century both for
     its conceptual importance and for its practical ramifications. It realised, in the hands of
      Rutherford, the alchemist's dream of the transmutation of elements and gave birth to the new
      disciplines of nuclear physics and elementary particle physics, which in turn revolutionised
      the understanding of astrophysics. Indeed, aleady in 1903 in a paper on radioactivity by
      Rutherford and Soddy~\cite{RS} can be found the prophetic statement:
      \par {\tt All these considerations point to the conclusion that the energy latent 
        in the atom must be enormous compared with that rendered free in ordinary \newline chemical
     change. Now the radio elements differ in no way from the other \newline elements in
    their chemical and physical behaviour. On the one hand they \newline resemble chemically their
    inactive prototypes in the periodic table very  \newline closely and on the other they possess
    no common chemical characteristics \newline that could be associated with their radioactivity.
    Hence there is no \newline reason to assume that this enormous store of energy is possessed by
    the \newline radio elements alone. It seems probable that atomic energy in general is\newline of 
   a similar high order of magnitude, although the absence of change \newline prevents its existence 
   being manifested.}~\footnote{This passage
    was quoted by Pais~\cite{PaisRMP77} who remarked that, in it, the modern concept of {\it atomic energy}
    was first introduced.}
   \par In this passage Rutherford and Soddy conjecture, on the basis of experimental evidence, 
     the universal nature of mass/energy equivalence, as later derived by Einstein in the first 1905 special
    relativity paper~\cite{Ein1}, as well as the necessity of transmutation, as considered in the
    second one~\cite{Ein2}, in order demonstrate its existence. This was just what was suggested
    by Einstein in the passage from Ref.~\cite{Ein2} quoted above. Indeed, as pointed out by Fadner~\cite{Fadner},
    the equivalence of mass and energy in radioactive decays, as suggested by Einstein, to test Eq.~(2.8) in
    Ref.~\cite{Ein2} had previously been conjectured, in an experimental context, by Soddy in a book published
    the previous year~\cite{Soddy}:
   \par {\tt ...it is not to be expected that the law of conservation of mass will \newline hold true for
     radioactive phenomena. The work of Kaufmann may be taken as an  experimental proof of the
    increase of apparent mass of the electron when \newline its speed approaches that of light.
   Since during disintegration
   electrons are expelled at speeds very near to that of light, which, after expulsion, experience
  resistance and suffer diminution of velocity, the total mass must be less after disintegration than
   before. On this view, atomic mass must be regarded as a function of the internal energy, and the dissipation
   of the \newline latter occurs at the expense, to some extent at least, of the mass of the \newline system.}
   \par       Einstein might also
      have noticed the similarity between (4.7) and (2.3) in order to write the relation
    \begin{equation} 
     E'(L,\beta) \equiv \gamma m(L) c^2
    \end{equation}
     where $m(L)$ is the `effective mass' of the radiated light. In this case the energy
     conservation equation (2.1) becomes, in virtue of (4.7) and (4.8)
    \begin{equation} 
   m_i c^2 = m_f c^2 +m(L) c^2
    \end{equation}  
    which is the conservation law of mass in Newtonian mechanics, also valid for static
    systems in relativistic mechanics. Consideration of (4.9) and the corresponding equation, (2.2),
    in the moving frame:
     \begin{equation} 
   \gamma m_i c^2 = \gamma m_f c^2  + \gamma m(L) c^2
      \end{equation}
     shows that Eq.~(2.8) may be derived without any necessity to consider the transformation 
      formula for the energy of `light waves' since the energies of the objects $i$ and $f$
     and the radiation $L$ all have the same transformation law (4.7) that was already
     implicit in the general relation (4.1) first given in Ref.~\cite{Ein1}. As clearly and correctly stated by Einstein,
      the relation (2.8) is indeed
     of complete generality, the specialisation to radiated energy in the thought experiment 
    being of no importance.
    \par Of particular practical interest are Eqs.~(4.8) and (4.9) for the special case
     $m_f = 0$ so that the entire energy content of the object $i$ is constituted by the
     two `plane waves' that are created ---photons in modern parlance. Measurement of the total energy of
    the photons in the rest frame of $i$ then enables the mass of the object to be determined.
    It was by such measurements that the neutral pi-meson was discovered in 1950
   by observation of the decay mode: $\pi^0 \rightarrow \gamma \gamma$~\cite{PI0Disc}.
    More recently, observation of the decay: ${\rm H} \rightarrow \gamma \gamma$ of the Higgs boson (H)
    made an important contribution to the discovery of this particle~\cite{CerncHB,AtlasHB,CMSHB}.
    \par    \par The formula for the transformation law of a `light complex' or `plane wave' derived in 
    Ref.~\cite{Ein1} and assumed in Ref.~\cite{Ein2}:
     \begin{equation}
      E'(\pm) = E\gamma(1\pm\beta \cos \phi),
     \end{equation} 
    where $\phi$ is the angle between the direction of one of the `plane waves' and the direction 
    of motion in the rest frame of the object $i$, was used to derive Eq.~(2.3) above as
     $E'(L) = E'(+)+E'(-)$.  
    It is the same, as Einstein remarked
    in Ref.~\cite{Ein1}, to the transformation law of the
  frequency, $\nu$, of the light wave. Perhaps surprisingly, he did not notice  ---or if he noticed, chose not to
   say--- that the identity
   of the transformations is a necessary consequence of the Planck relation $E = h \nu$, and although
   he had published, earlier in the same year, the paper~\cite{Ein3} in which the light quantum concept
   was introduced, did not then identify the `light complex' of Ref.~\cite{Ein1} with a light quantum or a group
   of light quanta in parallel motion.
   In fact, identical transformation laws for energy and frequency require that the ratio
    $E/\nu$ is the same in all inertial frames, and gives an alternative way to introduce Planck's
    constant, $h$, into physics~\cite{JHFEJP}. For further discussion of the possible role of
    Einstein's light quantum concept in the genesis of Ref.~\cite{Ein1} see Ref.~\cite{Ryansiewicz}.
    \par It is interesting to note that if Einstein had considered another application of Newton's Second Law,
    where force is equated to the time derivative of momentum, a straightforward variation of
    the kinetic energy calculation of Eq.~(4.6) from Ref.~\cite{Ein1} leads directly to the
    formula for relativistic momentum given later by Planck~\cite{Planck}:
  \begin{equation}
   p = \int dp = \int F dt = m\int_0^v\gamma^3\frac{dv}{dt} dt = mc\int_0^{\beta}
    \frac{d\beta}{(1-\beta^2)^{\frac{3}{2}}} = \frac{mv}{\sqrt{1-\beta^2}}.            
     \end{equation}
     \par Combining the relativistic energy: $E = \gamma m c^2$, momentum: $p = \gamma m v$ and Eq.~(3.10)
 relating $E$, $p$ and $m$ gives $v = pc^2/E = p c^2/\sqrt{m^2 c^4+p^2 c^2}$ so that for a massless particle
    $v = c$ and $E = p c$. Einstein could, in this way, at any time after 1906, have derived the second
    postulate of special relativity ---the constancy of the speed of light--- by assuming that the light
    quanta that he proposed~\cite{Ein3} in 1905 were massless particles~\cite{JHFHPA}.
\SECTION{\bf{Discussion}}
 Einstein's derivation of the equivalence of mass and energy in Ref.~\cite{Ein2}, expressed mathematically
  in Eq.~(2.8) above, was questioned by Planck in 1907~\cite{Planck1907}, Ives in 1952~\cite{Ives} and more
  recently by Ohanian~\cite{Ohanian} and Hecht~\cite{Hecht}. 
   \par Planck's objection, discussed by Fadner~\cite{Fadner}, was that, as a consequence of a possible
   contribution of thermal radiation to the rest energy of a system, the latter would be frame-invariant
   only to first order in $v/c$. However the total momentum of thermal radiation in the rest system
  vanishes, so that just as in the case of the two-photon system considered in Section 5 above, the
  thermal radiation component, $E_0^{\rm rad}$ of the rest energy has the same transformation
   law: $E^{\rm rad} = \gamma E_0^{\rm rad}$ as a ponderable object of mass $E_0^{\rm rad}/c^2$, an
   exact formula that is valid at all orders in  $v/c$. 
   \par In the paper `Derivation of the Mass-Energy Relation'~\cite{Ives} Ives reproduced Einstein's
     thought experiment of Ref.~\cite{Ein2} replacing the energy conservation postulate with relativistic
    momentum conservation. He then accused Einstein of the logical error of {\it petitio princpii}
    (introducing a premise logically equivalent to the claimed conclusion). The last sentence of the paper
   is the bald statement:
   \par {\tt The relation $E = m_M c^2$ was not derived by Einstein.}
    \par Here `$m_M$' is the mass of `matter' as opposed to the effective mass of electromagnetic
     radiation `$m_R$' also considered in the paper. This conclusion was subsequently quoted in
    an uncritical manner in several textbooks~\cite{Miller,Jammer,Arzelies}.
     \par Ives's thought experiment was identical to that considered by Einstein, except that the equal pulses
    of radiation of energy $E/2$ in the rest frame of the object $i$,
    were assumed to be emitted parallel and anti-parallel to the direction of
    motion of the objects $i$ and $f$. The energy of the pulses, in the frame where the objects are in
    motion with speed $\beta c$, using the same transformation formula, (4.11), for the energy of a
    `light complex' as that derived by Einstein in Ref.~\cite{Ein1} and used in  Ref.~\cite{Ein2}
     are:
      \[(E/2)\gamma(1+\beta),~~~~~ (E/2)\gamma(1-\beta). \] 
     as also given by Ives in Ref.~\cite{Ives}. 
      Ives then invoked a relation given by Poincar\'{e} in 1900~\cite{Poincare1900} stating that
       the momentum of electromagnetic radiation in free space is $S/c^2$ where $S$ is the 
     energy flux. Further writing $S = Ec$, Ives obtained for the (oppositely directed) momenta
     of the radiation pulses:  
      \[[E/(2c)]\gamma(1+\beta),~~~~~ [E/(2c)]\gamma(1-\beta) \]
     so that the net momentum of the radiation in the frame in which $i$ and $f$ are in motion
     is $E\gamma \beta/c$. Ives then invokes the formula (unknown to Einstein at the time of
     writing Ref.~\cite{Ein2}) for the relativistic momentum of a ponderable object of
      rest mass $m$ and velocity $\beta c$: $p = \gamma \beta mc$. Imposing conservation
     of relativistic momentum:
      \[ \gamma \beta m_i c = \gamma \beta m_f c + E\gamma \beta/c \] 
     immediately yields Eq.~(2.8) above: $ m_i - m_f = E/c^2$, the mathematical expression of
     Einstein's verbal conclusion, concerning mass/energy equivalence, in Ref.~\cite{Ein2}.
     \par Ives therefore obtains exactly the same result as Einstein in Ref.~\cite{Ein2} but assumes
     in addition to the transformation law of radiant energy, also the definition of
     the relativistic momentum of a ponderable object as well as Poincar\'{e}'s relation 
     between the energy flux and momentum of electromagnetic radiation. Instead, like Einstein
     in Ref.~\cite{Ein2}, of imposing energy conservation and the validity of Newtonian kinematics
     in the $\beta \rightarrow 0$ limit (both very weak postulates) Ives' derivation thus requires
     two additional, strong, postulates of relativistic physics.
     \par In his criticism of Einstein's analysis of the thought experiment Ives, unlike Einstein,
       invokes the relativistic formula for the kinetic energy of a ponderable object from
       Ref.~\cite{Ein1}:
     \begin{eqnarray}
     K_i & = & m_i\left[\frac{1}{\sqrt{1-\beta^2}}-1\right], \\
     K_f & = & m_f\left[\frac{1}{\sqrt{1-\beta^2}}-1\right].
    \end{eqnarray}
       As pointed out above, combining these equations with Einstein's exact relation (2.4), from 
       Ref.~\cite{Ein2},
      enables derivation of (4.4) from which follows in all generality ---without consideration
      of the $\beta \rightarrow 0$ limit--- the relation $m_i-m_f = E(L)/c^2$. By false logic,
      Ives construed the {\it general derivation} of Eq.~(2.8) just presented as a
       {\it petitio principii}. To do this he combined (5.1), (5.2) and (2.4) to obtain:
      \begin{equation}
         E'_i-E_i-(E'_f-E_f) = \frac{E(L)}{(m_i-m_f)c^2}(K_i-K_f)  
      \end{equation}
        which, Ives stated, follows necessarily from the equations:
     \begin{eqnarray}
     E'_i-E_i & = &  \frac{E(L)}{(m_i-m_f)c^2}(K_i+C), \\
     E'_f-E_f & = &  \frac{E(L)}{(m_i-m_f)c^2}(K_f+C).
    \end{eqnarray}
     These equations are consistent with Einstein's definitions of kinetic 
    energy if, and only if, $E(L)/[(m_i-m_f)c^2] = 1$. Ives then concluded that,
   in writing (2.5) and (2.6), as the definitions of kinetic energy, Einstein must 
   have (implicitly) assumed that $E(L)/[(m_i-m_f)c^2] = 1$ which is the result
   he claimed to prove! But, as pointed out in Section 4 above, the relation 
   $E(m,\beta = 0) = mc^2$ (which Ives claims Einstein did not derive) is a consequence
   of (5.1) or (5.2) {\it alone}. It is also a consequence of (2.8) {\it alone} since on setting 
    $m_f = 0$ in this equation it is found that:
    \begin{equation}
    m_i c^2 = \Delta E_i = E(m_i, \beta = 0) = E(L).
    \end{equation}
     It is also possible to derive, in a similar manner, Eq.~(4.1) from equations (2.4)-(2.6) 
    and (2.8) from Ref.~\cite{Ein2}. Combining these equations gives:
        \begin{equation}
          K_i-K_f = (m_i-m_f)c^2\left[\frac{1}{\sqrt{1-\beta^2}}-1\right].  
        \end{equation}
      Eq.~(4.1) follows on setting $K_f = 0$, $m_f = 0$.      
    \par  Unlike Ives, Einstein {\it did not invoke} the relativistic kinetic energy
     formula (4.1) in Ref.~\cite{Ein2}. The relation $E = m_M c^2$ mentioned in the last
    sentence of Ref.~\cite{Ives}, quoted above, which, in the notation of the present
    paper, is $E(m,\beta = 0) = mc^2$, can be derived {\it either} from (2.8) (without invoking (4.1))
    which is what is done in Ref.~\cite{Ein2} {\it or},  from (4.1) in
    Ref.~\cite{Ein1} (without invoking (2.8))
    ---which can be done by simply physically interpreting
    this equation--- although Einstein did not do this in
    Ref.~\cite{Ein1}.
    In neither case did Einstein introduce a premise
    logically equivalent to a result he claimed to derive. For further critical discussion
    of Ives' assertion that Einstein was guilty of {\it petitio principii} in Ref.~\cite{Ein2}
     see Refs.~\cite{ST,Fadner}.
    \par In a recent paper~\cite{Ohanian} Ohanian has claimed that Einstein's derivation of
    $E_0 = mc^2$ in Ref.~\cite{Ein2} is flawed by the assumption that the kinetic energy, $K$,
    of an extended body is given by the Newtonian formula: $K = mv^2/2$ in the low-velocity
     limit. It is conjectured that the Newtonian separation of internal energy, contributing
   to the rest mass $m$, and translational kinetic energy $K$, may break down in the case that
   internal constituents of the body are in motion with relativistic velocities. To be specific,
   the approximate formula for the kinetic energy as (tacitly) assumed by Einstein:
   \begin{equation}
     K(v) = E-E_0 = \frac{1}{2} m v^2
   \end{equation}
    is conjectured by Ohanian to be modifed to
   \begin{equation}
    \tilde{K}(v) = \tilde{E}-E_0 = \frac{1}{2}\tilde{m}(v) v^2
   \end{equation}
    where $\tilde{m}(v) \ne m$ when $v \ne 0$ in the case that constituents of a system with rest mass
   $m$ have relativistic velocities.
     \par A Taylor expansion of  $\tilde{m}(v)$ gives:
      \begin{equation}
       \tilde{m}(v) = m+ \tilde{m}(v)'v+ \frac{ \tilde{m}(v)''}{2!}v^2 +.~.~.
      \end{equation}
     since, by the definition of rest mass, $ \tilde{m}(0) = m$. Defining rest masses, 
     following Stachel and Torreti~\cite{ST} as:
       \begin{eqnarray}
       m  & \equiv & \left[Lim~v \rightarrow 0\right] \frac{K(v)}{v^2/2}, \\
     \tilde{m}_0  & \equiv & \left[Lim~v \rightarrow 0\right] \frac{\tilde{K}(v)}{v^2/2}
      \end{eqnarray}
      it follows from (5.9) and (5.10) that $\tilde{m}_0 = m$ so that modifications of the Newtonian
      kinetic energy (5.8) as suggested by Ohanian cannot invalidate the derivation of $E_0 = mc^2$
      given in Ref.~\cite{Ein2}. The reason for this is that both the rest mass $m$ and $E_0$ in 
    the formula $E_0 = mc^2$ are, by definition, {\it static properties} of a ponderable body, for which
    there is no difference between Newtonian and special relativistic kinematics.
   \par In Ref.~\cite{Ohanian}, Ohanian recalls and demonstrates a proof due to Klein~\cite{Klein}
    of the four-vector character of the quantities: $\int T^{\nu 0}d^3x$ where $T^{\nu 0}$
    are elements of the energy-momentum tensor of an arbitary extended and closed physical system.
    It is claimed in Ref.~\cite{Ohanian} that a corollary of this proof is that:
     \par {\tt For any closed system, with a timelike energy momentum four-vector the \newline 
           energy $E_0$ in the zero momentum frame ``rest frame'' is related to the mass
       by $E_0 = mc^2$.}
    \par To show this Ohanian then considers a Lorentz transformation of the energy momentum four-vector:
     $(E_0/c,0,0,0)$ to give the energy and momentum in an arbitary inertial frame:
     \begin{eqnarray}
      E & = & \gamma E_0,  \\
      \vec{p} & = & \frac{\gamma E_0 \vec{v}}{c^2}.
     \end{eqnarray}
     Eq.~(5.13) gives, for the relativistic kinetic energy: $K = \gamma E_0 - E_0$, Comparing 
    this with the relativistic kinetic energy formula $K = mc^2 \gamma - mc^2$   given
    by Einstein in Ref.~\cite{Ein1} Ohanian claims to have then derived the formula  $E_0 = mc^2$
    as a rigorous consequence of the four-vector character of certain elements of the 
    energy momentum tensor, as proved by Klein. However, since Ohanian also assumes
   the correctness of (4.6):
  \[   K = mc^2 \gamma - mc^2 \]
    or, defining verbally the terms in this equation:
  \[ {\rm kinetic~energy}(K)  = {\rm energy~in~motion}(E) - {\rm rest~energy}(E_0) \]
   the relation  $E_0 = mc^2$, as already pointed out in Section 4 above, is a consequence
   of Eq.~(4.6) alone and the definitions of  kinetic~energy, energy~in~motion and rest energy.
    Klein's proof is therefore irrelevant for the derivation of $E_0 = mc^2$ ---it follows
    directly from Eq.~(4.6) which Ohanian introduces as a separate premise in his proof.
    Unlike Einstein in Ref.~\cite{Ein2}, as claimed by Ives, Ohanian is indeed guilty of 
     {\it petitio principii} in Ref.~\cite{Ohanian}!
     \par The abstract of a recent paper `How Einstein confirmed $E_0 = mc^2$'~\cite{Hecht} by Hecht
     concludes with the assertion: 
    \par {\tt Although he repeatedly confirmed the efficacy of $E_0 = mc^2$, he never \newline constructed 
     a general proof. Leaving aside that it continues to be affirmed experimentally, a rigorous
      proof of the mass-energy equivalence is probably beyond the purview of the special
     theory.} 
    \par  A detailed reading of Ref.~\cite{Hecht} did not reveal to the present author
     what type of `proof' of mass-energy 
     equivalence Hecht would consider to be `rigorous'. Presumably not, in view if the above
     quotation, the experimental proof that it is indeed observed to be an important property
    of the real world. In mathematics, a `proof' is the demonstration that a certain assertion
     is a logical consequence of certain stated premises. Whether the premises describe, exactly
     or approximately, some measureable feature of the real world is irrelevant to the correctness
     or level of rigour of such a proof. On the other hand, physics is only concerned with 
     premises that do describe, exactly or approximately, some measureable feature of the real world.
     Einstein's proof of mass-energy equivalence in Ref.~\cite{Ein2} is based on just three 
    premises: conservation of energy, the transformation law of electromagnetic energy, and the
   assumption that in the low velocity limit, ($v \ll c$), the laws of Newtonian mechanics are valid.
   It should be noticed that all of these premises {\it do} correctly describe observed features of the
   real world and so are physically valid ones. Unless there is some logical mistake in the mathematical
    reasoning (essentially only algebra) leading to Einstein's conclusion (2.8) ---the exact 
   mathematical expression of his verbal statement of mass-energy equivalence--- the proof certainly
   follows logically from the premises. No convincing argument is given by Hecht in support of
   his assertion that it is `not rigorous'.
    \par One objection raised by Hecht is that: 
    \par {\tt At this time he could imagine that there might be some kind of quiescent matter that
      possessed residual inert mass even if all its energy was \newline somehow removed. After all, light
    was an entity with energy and no mass; \newline perhaps there was matter with mass and no energy.}
     \par Since setting $m_f = 0$ in Eq.~(2.8) gives $m_i = E_i(v = 0) c^2$ not
   $m_i+ \mu_i = E_i(v = 0) c^2$ where the symbol $\mu_i$ represents Hecht's
 `matter with mass and no energy' the existence of such `energyless mass' with $\mu_i \ne 0$ is then 
   forbidden 
     if Eq.~(2.8) is correct. Such mass can therefore exist only if one of the
    three premises on which  Eq.~(2.8) is based is false. Hecht, following Ohanian, suggests that it may be the 
    last premise (made tacitly by Einstein) that is false, i.e. that there is
    a breakdown of Newtonian mechanics at low velocities. As pointed out above, this objection
    is rebutted since the equation $E_0 = m c^2$ describes a {\it static} system so that the precise 
    definition of `kinetic energy' used to derive Eq.~(2.8) is irrelevant. Indeed, from dimensional
   analysis alone, the relation between rest mass $m$ and the corresponding `rest energy' $E_0$
    must be of the form $E_0 = \kappa m$ where $\kappa$ is a universal constant with the
    dimensions of velocity squared. Then, necessarily, $m = 0$ when $E_0 = 0$ and vice versa.
     The existence of matter in motion with `energy and no mass' is a known special feature
     of relativistic kinematics. Taking the simultaneous limits: $m \rightarrow 0$,
     $ \beta \rightarrow 1$ in the formula (3.6) for relativistic energy gives, formally:
     $0 \times \infty = E = pc$, as is clear from from inspection of the formula (3.10)
     given by squaring both sides of Planck's formula (3.8) relating the 
    rest mass of an object to its relativistic energy and momentum.
   \par As discussed in Section 4 above, Hecht's assertion in Ref.~\cite{Hecht} that:
   `Einstein said nothing about' ($E_0 = mc^2$) `in his June 1905 paper.'(\cite{Ein1}), while
    perhaps literally true, is highly misleading as to the actual scientific content
    of this paper. If `said' is interpreted only as verbal expression, the statement is
    correct. However, the relation $E_0 = mc^2$ follows trivially from Eq.~(4.7), which is simply
    a transposition of Einstein's equation (4.1) from Ref.~\cite{Ein1}, in association
    with the definition of relativistic energy: $E \equiv \gamma
   mc^2$. The manner of stating  mass-energy
    equivalence is inverted in the June 1905 paper~\cite{Ein1},where Eq.~(4.1) is given without
    verbal explanation, as compared to the September 1905 paper~\cite{Ein2} where an exact verbal
    description of this equivalence is given, but not the corresponding equation (2.8).
    \par Hecht also states, \`{a} propos of the September 1905 paper that:
    \par {\tt This derivation came very early in the development of relativity and the formal 
     concept of ``rest energy'' had not yet evolved, nor had $E_0$ been \newline introduced to
     symbolize it.}
     \par  This statement is untrue. The quantities $E_i$ and $E_f$ in Eq.~(2.1)
      (called by Einstein $E_0$ and $E_1$ in Ref.~\cite{Ein2}) are, by definition, the
       {\it rest energies} of the objects $i$ and $f$! Thus Einstein
   {\it did} use the symbol $E_0$ to denote the energy of the object
   $i$ when it is at rest. The formula in Ref.~\cite{Ein2}
      corresponding to Eq.~(2.1) is not mentioned in Ref.~\cite{Hecht}. Also not mentioned
    is Einstein's verbal statement of the exact conversion factor between mass and
    energy given in the concluding passage of Ref.~\cite{Ein2} quoted above. Hecht 
    states instead that:
      \par {\tt Nowhere did he write that mass and energy are ``equivalent'', that would 
      come later.}
    \par However Einstein did state that one gramme of matter corresponds to $9\times 10^{20}$
      ergs of energy, and to say that one gramme is `equivalent' to $9\times 10^{20}$
      ergs does not have a different meaning. Indeed, in Ref.~\cite{Hecht} Hecht does not 
     consider several written passages or equations in
   Ref.~\cite{Ein2} that actually contradict some assertions
    made in the Ref.~\cite{Hecht}.
    \par The conclusion of Ref.~\cite{Hecht} also contains the statment (not directly related to the
    meaning of Einstein's work in Refs.~\cite{Ein2,Ein1}): 
   \par {\tt For several compelling reasons many physicists have come to accept that mass
     is invariant, even though there is no proof ---theoretical or \newline experimental--- that it is.}
     \par The general `theoretical proof' that rest mass is invariant is provided
    by the straightforward generalisation of Eq.~(4.8) above, where the `effective mass' of
    a two-photon system is introduced, to an arbitary 
    final state of
    $N$ ($N \ge 2$) particles.
     One `experimental proof' is provided by the observation of the Higgs boson, that is produced in
    the laboratory system at the LHC with a wide energy spectrum~\cite{AtlasHB,CMSHB}, via its
   two photon decay mode, as mentioned in Section 4 above. Another is the identification of 149
   distinct decay modes of the $J/\psi$ (a bound state of a charm quark and a charm anti-quark)
   ~\cite{PDG2012} by calculation of the Lorentz-invariant effective mass of the different
   decay products and observing that they are all equal to the mass of the $J/\psi$.
    The evidence, both theoretical and experimental, is indeed `compelling'. What
   kind of further `proof', that would convince Hecht that rest mass is indeed invariant and equivalent
   to energy, is not revealed in Ref.~\cite{Hecht}.     
 \SECTION{\bf{Conclusions}}
    The relation $E_0 = mc^2$ and the concept of total relativistic energy: $E = \gamma mc^2$ were already
    implicit in Eq.~(4.1) given in Einstein's June 1905 special relativity paper~\cite{Ein1}. The important
   new contribution to physics brought by the September 1905 paper~\cite{Ein2} was therefore, not only an alternative
   derivation of `$E_0 = mc^2$', but the first correct theoretical analysis of a process in which mass was
   transformed into energy and (in modern language) new particles were created from this energy. Experimental
   studies of radioactive decays had already shown by 1905 that such energy transformation processes must 
    occur in nature, and Einstein suggested that such radioactive decays might be used to test his theory.
    Planck wrote down in Ref.~\cite{Planck} the formula for relativistic momentum:
    $p = \gamma m v$, and by invoking Hamilton's Principle, rederived Einstein's relativistic energy
    equation (4.1), thus completing the theory of the relativistic kinematics of objects (massive or massless)
    in motion in free space. The formula for relativistic momentum can alternatively be derived by time integration,
    as in Eq.~(4.11), of the relativistic generalisation of Newton's Second Law: Eq.~(4.5), as
     given in Ref.~\cite{Ein1}.
   \par In agreement with previous work of Stachel and Torreti~\cite{ST} and Fadner~\cite{Fadner}, claims in the
   literature that Einstein's discovery of mass-energy equivalence, as presented in Ref.~\cite{Ein2}, was flawed, 
   incomplete, or lacked generality or rigour~\cite{Planck1907,Ives,Ohanian,Hecht} are shown to have no
    foundation.
  
\end{document}